\documentclass[10pt, twocolumn, conference]{IEEEtran}
\usepackage[utf8]{inputenc}
\usepackage[perpage,symbol]{footmisc}
\usepackage[english]{babel}
\usepackage{multirow,multicol}
\usepackage{amsmath, bm} 
\usepackage{pbox, booktabs, rotating}
\usepackage{subfigure}
\usepackage{array, makecell, lipsum}
\usepackage{enumerate}
\usepackage{threeparttable}
\usepackage{multirow,hyphenat,balance}
\usepackage{placeins}
\usepackage[compress]{cite}
\usepackage{float,color}
\usepackage{ragged2e}
\usepackage{enumerate}
\usepackage{etoolbox}
\usepackage[linesnumbered,ruled]{algorithm2e}
\usepackage{paralist}
\usepackage{cases}
\usepackage{url}
\usepackage{framed}
\usepackage{flushend}
\usepackage{amsmath}
\usepackage{amsthm}

\usepackage{amsmath}
\usepackage{amssymb}

\usepackage{caption}
\theoremstyle{definition}

\newcolumntype{M}[1]{>{\centering\arraybackslash}m{#1}}

\newcommand*{\ie}{{\em i.e.,}\@\xspace}
\newcommand*{\eg}{{\em e.g.,}\@\xspace}

\title{A Novel Tampering Attack on AES Cores with \\ Hardware Trojans} 
\author{\IEEEauthorblockN{Ayush Jain, and Ujjwal Guin \\}
\IEEEauthorblockA{Dept. of Electrical and Computer Engineering, Auburn University\\
Emails: \{ayush.jain, ujjwal.guin\}@auburn.edu}
}

\begin{document}

\maketitle
\thispagestyle{plain}
\pagestyle{plain}

\begin{abstract}
The implementation of cryptographic primitives in integrated circuits~(ICs) continues to increase over the years due to the recent advancement of semiconductor manufacturing and reduction of cost per transistors. The hardware implementation makes cryptographic operations faster and more energy-efficient. However, various hardware attacks have been proposed aiming to extract the secret key in order to undermine the security of these primitives. In this paper, we focus on the widely used advanced encryption standard~(AES) block cipher and demonstrate its vulnerability against tampering attack. Our proposed attack relies on implanting a hardware Trojan in the netlist by an untrusted foundry, which can design and implement such a Trojan as it has access to the design layout and mask information. The hardware Trojan's activation modifies a particular round's input data by preventing the effect of all previous rounds' key-dependent computation. We propose to use a sequential hardware Trojan to deliver the payload at the input of an internal round for achieving this modification of data. All the internal subkeys, and finally, the secret key can be computed from the observed ciphertext once the Trojan is activated. We implement our proposed tampering attack with a sequential hardware Trojan inserted into a 128-bit AES design from OpenCores benchmark suite and report the area overhead to demonstrate the feasibility of the proposed tampering attack.    
\end{abstract}

\vspace{5px}

\section{Introduction} \label{sec:intro} 
The advancement in semiconductor manufacturing and testing has enabled the system-on-chip~(SoC) design house to incorporate more functionality in modern SoCs that consists of millions of transistors. Consequently, the overall complexity of designing and manufacturing such integrated chips~(ICs) has increased. As advanced technology nodes are adopted, building and maintaining a foundry requires large capital investment~\cite{YehFabCost2012}, resulting in a minimal number of foundries across the globe. Currently, integration of third-party intellectual properties~(3PIP) with the original design and outsourcing to an offshore foundry for manufacturing and testing is a typical current trend for design-houses. However, this globalized and distributed supply chain model comes with ample scope to tamper the design by implanting hardware Trojans, malicious modifications into the IC, at the design and fabrication phases~\cite{tehranipoor2011introduction,adee2008hunt}. Hardware Trojans can pose a severe security threat to the designs used for security-sensitive applications, such as cryptographic modules.

Cryptographic algorithms have widely been adopted as a critical means to provide the security of communication, data, and other sensitive assets. The applications range from commonly used smart cards to highly critical defense or government applications, which rely on these algorithms implemented with dedicated hardware for confidentiality, end-point authentication, integrity verification, and non-repudiation~\cite{barker2016nist}. A secure hardware implementation of a cryptographic system comprises of key dependant logic operations, where the secret key is stored in a tamper-proof memory. The plaintext inputs of a system go through a series of cryptographic computations dependent on the secret key value to produce the final ciphertext. Traditionally, such hardware implementations should be secure even if all the design details, except the secret key, is publicly available. Unfortunately, a hardware Trojan can leak this secret key to an adversary once it is activated. 

The research community has been extensively studying the taxonomy of hardware Trojans, their implementation, and detection in cryptosystems (\ie AES, RSA, and ECC). These Trojans are designed to leak the secret key from the circuits, either through side-channels~\cite{lin2009trojan,lin2009moles,zhao2018novel} or primary outputs~\cite{kutzner2013hardware, becker2013stealthy, bhasin2013hardware, muehlberghuber2013red, bhunia2014hardware}.  Over the years, various detection and prevention methods have been proposed to address the threat originated from hardware Trojans. The detection methods can be categorized -- ($i$) logic testing~\cite{kitsos2015exciting, wang2008detecting, salmani2016cotd}, ($ii$) side-channel analysis~\cite{hanindhito2019hardware,liu2014hardware,he2017hardware,muehlberghuber2013red}. On the other hand, prevention methods include modifications in the design~\cite{wu2016aes,xiao2014novel,ba2016hardware} and split manufacturing~\cite{Vaidyanathan2014detecting, rajendran2013split, wang2016cat}. All of these techniques have some drawbacks in terms of their feasibility, the type of Trojans that can be targeted, and adverse effects due to manufacturing process variations. Besides, sequential Trojans manifest their effect only when a particular time has elapsed after the trigger condition is met or when the Trojan is triggered multiple times in a row. This property of sequential Trojan makes their detection very difficult. As a result, this type of Trojan can become a prominent and suitable choice for an attacker to launch tampering attacks.  

In this paper, we show how an adversary can extract the secret key from different implementations of Advanced Encryption Standard~(AES) by tampering the netlist with a hardware Trojan. AES performs a sequence of operations on the plaintext in multiple rounds that involves intermediate subkeys for each round, generated from the original secret key. Our attack relies on masking other intermediate subkeys' effect in an internal round through a sequential hardware Trojan. Once the Trojan is activated, it obstructs and modifies the data from all previous rounds. As a result, the input data for the Trojan-affected round becomes all 1s if the Trojan's payload is an OR gate. We can also use an AND gate as the payload to make the input data all 0s. We refer this as an adversarial known value~(\ie 0 or 1) because only the adversary pertains to the knowledge regarding this value and also the rare Trojan activation condition that would achieve this intentional alteration. The resultant output for this round can be observed directly from the primary output if a Trojan is implanted in the last round. 

The contributions of this paper are described as follows:

\begin{itemize}
    \item We propose a novel attack based on the malicious modifications of the hardware implementation of an AES core. The attack aims to modify the computation for an internal round and extract its corresponding intermediate subkey. For the same, we tamper the AES design with a sequential hardware Trojan. To the best of our knowledge, we are the first to demonstrate that the extraction of an intermediate key can be performed by inserting a sequential hardware Trojan, which can help an adversary for computing the original secret key. We propose to use the design for a sequential hardware Trojan due to its greater difficulty of detection during manufacturing tests and the normal functioning of the circuit. The addition of a state element~(a counter) to the trigger of sequential Trojan requires triggered Q times consecutively, to deliver the payload. 
    
    \item We demonstrate and validate our proposed attack on the OpenCores AES benchmark~\cite{opencores} synthesized in 32nm technology using Synopsys Design Compiler. The area and power overhead resulted from inserting a sequential hardware Trojan are negligible compared to the AES core. 
\end{itemize}

The rest of the paper is organized as follows. First, we describe the AES structure in Section~\ref{sec:background}. We present the proposed attack and its methodology on different AES implementations in Section~\ref{sec:proposed_attack}. Experimental results related to hardware Trojan and the proposed attack are shown in Section~\ref{sec:experimentation}. Finally, we conclude the paper and provide future directions in Section~\ref{sec:conclusion}. 

\section{Background} \label{sec:background}
Advanced Encryption Standard~(AES) is a widely used block cipher for data encryption recommended by the National Institute of Standards and Technology (NIST) in November 2001~\cite{standard2001announcing}. An adversary can tamper the AES core with a hardware Trojan as the implementation details of AES are publicly available.  In this section, we provide a detailed description of the AES core and a hardware Trojan, which can be used to launch the tampering attack described in Section~\ref{sec:proposed_attack}.   

\subsection{AES Block Cipher}
AES is the widely popular block cipher used almost in every secure application. It consists of multiple rounds of operations (\eg 10, 12, and 14) for different key sizes (\eg 128, 192, and 256, respectively). Each round~($R_i$ with $i\in\{1,2,\ldots, n\}$) consists of SubBytes~(SB), ShiftRows~(SR), MixColumns~(MC), and AddRoundKey~(AK) layers, except for final round without the MixColumns computation~\cite{standard2001announcing}. The intermediate round computations are usually represented by a 4$\times$4~matrix, where each cell represents a byte.  Note that the subscript for any variable represents the round number and superscript represents the accessible subgroups within that variable. The same notations are used and referred throughout the paper. We denote the input of the $i^{th}$ round by ${A_i}^j$, where $j \in \{0,1, \ldots, 15\}$. The internal round keys ($K_i$) are generated from the key expansion modules. These intermediate keys corresponding to each round can be denoted as subkeys. These subkeys are bitwise XORed with the output of the MixColumns (or ShiftRows for the final round). The key bytes are arranged into a matrix with 4 rows and 4~(128-bit key), 6~(192-bit key) or 8~(256-bit key) columns. In this paper, we only focus on AES with a 128-bit key to demonstrate the tampering attack for simplicity. This same attack can be launched for AES with 196 and 256-bit keys as well without changing the attack methodology.  

One can find the details for each round of AES in ~\cite{standard2001announcing} and can be summarized in different layers described as follows:

\begin{enumerate}
    \item SubBytes~(SB): It is the nonlinear transformation step in AES, where each state byte is swapped with a pre-computed value from a look-up table known as s-box.
    \item ShiftRows~(SR): This step rotates the $4\times4$~state matrix with different known offsets. Rows are shifted in a cyclic manner by 1, 2 and 3-bytes for the corresponding row number in the state matrix.
    \item MixColumnns~(MC): This step performs linear column-wise operations on the state matrix. Essentially, it is a matrix multiplication in the finite field of each column in the state matrix with a constant $4\times4$~matrix. 
    \item AddRoundKey~(AK): It is the bitwise XOR of the state matrix with the corresponding subkey.
\end{enumerate}

\begin{figure}[ht]
    \centering \vspace{-10px}
    \includegraphics[width=\linewidth]{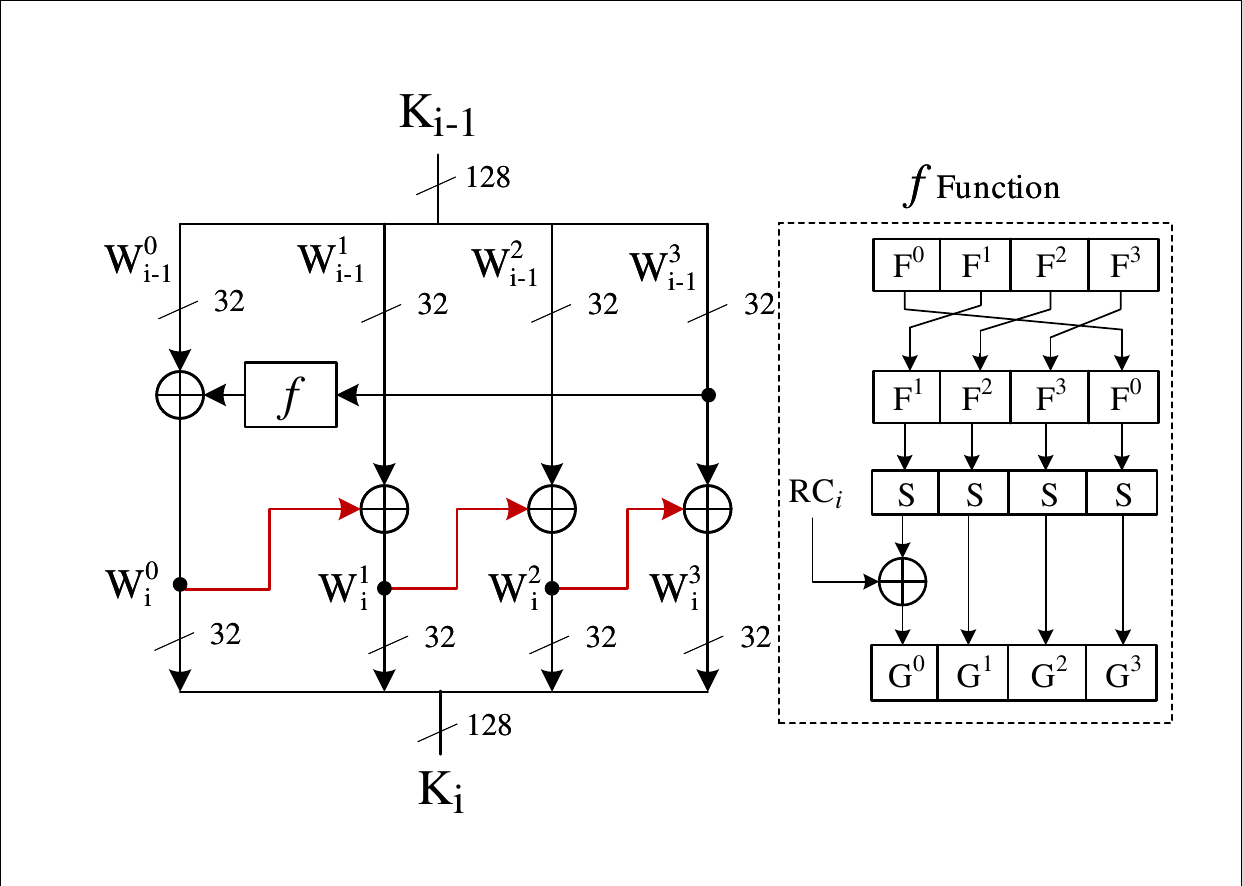} \vspace{-20px}
    \caption{Key schedule module for 128-bit AES implementation.}
    \label{fig:key_schedule}
\end{figure}

The key schedule (KS) module generates subkeys for each rounds. As the AES primitive with 128-bit key (can be referred as AES-128) has 10 rounds, it is necessary to create 10 subkeys~($K_1, \ldots, K_{10}$) of 128-bit each. Following the notations, original key for AES $K$ consisting of \{${k}^0,{k}^1, \ldots {k}^{15}$\}, where each subgroup comprises of 8-bits respectively. Figure~\ref{fig:key_schedule} shows the implementation details of the key schedule module for the $i^{th}$ round, where the current round's subkey ($K_i$) can be computed from the previous round's subkey ($K_{i-1}$). The subkey operations are performed on a word length~(32-bits or 4 bytes) subgroup. These subgroup words for each round can be represented as $W^r_i$, where $r$ and $i$ represent word index and round index, respectively. For $i^{th}$ round, $W^0_i$ is comprised of \{${k_i}^0,{k_i}^1,{k_i}^2,{k_i}^3 $\}. Similarly, $W^1_i$ is formed with \{${k_i}^4,{k_i}^5,{k_i}^6,{k_i}^7 $\} and so on. 

As shown in the Figure~\ref{fig:key_schedule}, we can generalise the operation for the key schedule module and be described as: 

\vspace{-10px}
\begin{align*}
    W^0_i&= W^0_{i-1} \oplus f(W^3_{i-1}) ; &W^1_i&= W^0_{i} \oplus W^1_{i-1}  \\
    W^2_i&= W^1_i \oplus W^2_{i-1}; &W^3_i&= W^2_i \oplus W^3_{i-1} 
\end{align*}
\vspace{-5px}
where, \textit{f} function can be formalised as:

\vspace{-10px}
\begin{align*} 
G^0 &= S(F^{1}) \oplus RC_i ;
&G^1 &= S(F^{2})  \\
G^2 &= S(F^{3})  ;
&G^3 &= S(F^{0})
\end{align*}

\vspace{-5px}
where, the values of $RC_i$ can be found in~\cite{standard2001announcing}. 

Note that the detailed implementation of the key schedule module will help to compute the previous round's subkey ($K_{i-1}$) and finally the secret key $K$, if any of the subkey ($K_i$) is known.  

\subsection{Design for a Sequential Hardware Trojan} \label{sec:sequential_Trojan}
\begin{figure}[t]
    \centering
    \includegraphics[width=0.7\linewidth]{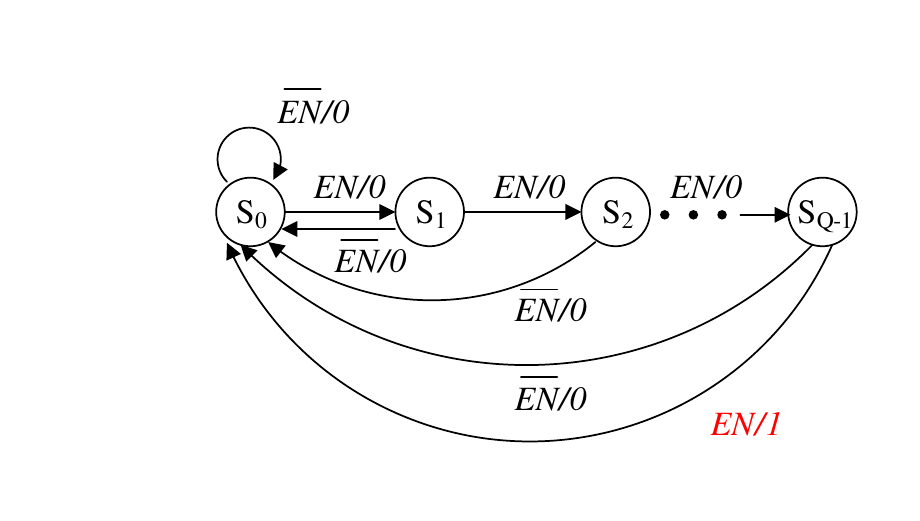}
    \caption{Finite state machine of the counter used in a sequential hardware Trojan.} \vspace{-10px}
    \label{fig:counter_FSM}
\end{figure}

In this paper, we consider the design of a sequential Trojan to demonstrate the attack. Upon triggering, a sequential Trojan manifests it effect after the occurrence of a sequence or a period of time. Generally, Trojan comprises of a trigger and payload that can be activated through trigger inputs, which are taken from the primary inputs and/or internal nodes of a circuit. The Trigger inputs are selected such that the Trojan can evade manufacturing or production test patterns~(\eg stuck-at fault tests, and delay tests)~\cite{zhou2018modeling,jain2019taal,jain2020special}. A \textit{Type-p} Trojan comprises of \textit{p} trigger inputs. The trigger is selected as an AND gate. However, any other combinational logic can also form the trigger which provides logic 1 when activated. Along with this AND gate, the sequential Trojan trigger includes a state element~(Q-State counter). Upon availability of trigger inputs, the output of this AND gate becomes 1 (\ie $EN=1$) and the counter is incremented by 1. Upon triggering the Trojan $Q$-times consecutively, the counter reaches the maximum value and delivers the payload in the original circuit through the OR gate or XOR gate. The finite state machine (FSM) for the counter~(CTR) is shown in Figure~\ref{fig:counter_FSM}. The state transition occurs only when $EN=1$, otherwise, it returns to the initial state, $S_0$. The output of the counter becomes 1, once $EN$ is made to logic 1 consecutively for $Q$ clock cycles. An adversary may choose any different design of a Trojan as well. 

\section{Proposed Tampering Attack on AES with a Hardware Trojan} \label{sec:proposed_attack}

The important aspect of cryptographic primitives is to encrypt the output in such a way that an adversary cannot find any key information at the output. In other words, no key information is leaked at the output and an adversary cannot determine the key by observing input/output responses of a system. In this section, we show how an adversary can extract the secret key using the proposed tampering attack with a sequential hardware Trojan.

\subsection{Threat Model}
The threat model is described to clearly identify the capabilities of an adversary. In this model, an untrusted foundry is considered as an adversary with the following capabilities: 

\begin{itemize}
    \item It has access to the netlist of the crypto primitives. The untrusted foundry has access to all the layout and mask information, which can be obtained from the GDSII or OASIS file. The netlist can be reconstructed from this information using reverse engineering~\cite{torrance2009state}.
    
    \item The attacker has the capability to modify the netlist so that it can tamper it with a hardware Trojan. 
    
    \item The attacker has access to all the manufacturing test (\eg stuck-at fault and delay fault) patterns as it is common that production tests are performed at the foundry. The adversary can utilize these test patterns to design a Trojan which cannot be detected during manufacturing tests~\cite{zhou2018modeling}. 
    
\end{itemize} 

\subsection{Attack Methodology}

The proposed attack relies on tampering the netlist by an untrusted foundry with the aim of exposing the secret key. Once the secret key is exposed, the security of AES no longer exists. With this aim, an efficient two-step methodology is proposed that involves a sequential Trojan. The proposed attack can be described as follows:
\begin{itemize}
    \item \textit{Step 1}: The first step is to implement the hardware Trojan and place its payload in the netlist. Once activated, the Trojan masks the information obtained from previous round computations and nullify the impact of previous subkeys. The AddRoundKey~(AK) layer is our primary area of interest while tampering the circuit with a hardware Trojan and modify intermediate round state matrix~($A_i$) to all 1s or 0s depending on the payload. In this paper, we treat the payload as OR gates, and thus $A_i$ becomes all 1s. Once the response is collected at the primary output, an adversary then computes the secret key using \textit{Step 2}.

    \item \textit{Step 2}: The computation of the secret key is performed in this step from the Trojan activated response. As the design for the key schedule module is publicly available, an adversary can compute the input data of the key schedule module from its output (see the computation details in Algorithm~\ref{alg:inv_key_schedule} presented in Section~\ref{subsec:AES}). With this, any subkey can be traced back to retrieve the previous round key~($K_{i-1}$) and finally, the original secret key~($K$).
\end{itemize} 

\subsection{Tampering Attack on AES Core} \label{subsec:AES}

\begin{figure}[t]
    \centering
    \includegraphics{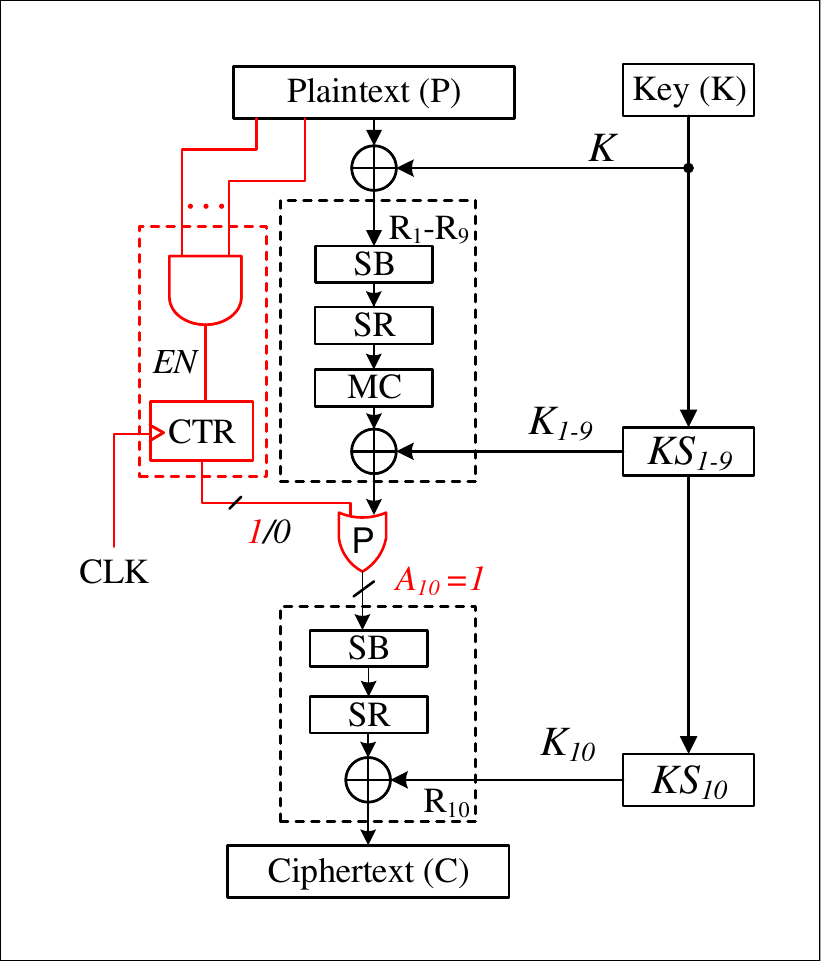} \vspace{-5px}
    \caption{Tampering attack on AES design implanted with a sequential hardware Trojan, delivering the payload at the data input~($A_{10}$) of the last round~($R_{10}$).} \vspace{-10px}
    \label{fig:Attack_AES}
\end{figure}

Figure~\ref{fig:Attack_AES} shows the proposed tampering attack on 128-bit the AES implementation. The netlist is implanted with \textit{Type-p} sequential hardware Trojan, which consists of \textit{p}-input trigger and a payload OR gate. The trigger inputs may come from the input depicted as plaintext ($P$) in the figure. The trigger circuit comprises of a AND gate and counter~(CTR) with the maximum count number of \textit{Q}. One can find the detailed description of this sequential Trojan in Section~\ref{sec:sequential_Trojan}. We propose to place the payload OR gate(s) before the last round~($R_{10}$) computation, \ie after the add round key, \textit{AK} layer in $R_9$. Once the Trojan is activated, the output of the payload~(P) OR gate(s) becomes 1. As a result, all the bits in the input state matrix for last round $A_{10}=1$, consequently, masking the effect of all previous round key information~($K_1-K_9$) and round computations~($R_1-R_9$). Due to availability of the implementation details, we can compute the $K_{10}$ from the observed ciphertext $C$, and described as follows: 

\begin{enumerate}
    \item The activated Trojan delivers the payload to modify the input to the last round with $A_{10}= 0\text{x}FF\ldots FF$. 

    \item The output of SubBytes~(SB) layer of $R_{10}$ can be computed as: 
        \vspace{-10px}
        \begin{align*}
        Y_1&=SB(A_{10}) \\
           &= 0\text{x}1616\ldots1616 
        \end{align*}
        
    \item Once $Y_1$ is known, the output of ShiftRows~(SR) layer can be computed as: \vspace{-5px}
    \begin{align*}
    Y_2&=SR(Y_1) \\
    &= 0\text{x}1616\ldots161616
    \end{align*}
\end{enumerate}

Since the last round $R_{10}$ does not perform MixColumn~(MC) operation, the output of ShiftRow~(SR) gets XORed with subkey~($K_{10}$) in AddRoundKey~(AK) step which is the final ciphertext~(C) at the primary output. From this output bitwise XOR being a symmetric operation, we can calculate the subkey~($K_{10}$) as: \vspace{-20px}

\begin{align*}
    K_{10} &= C \oplus Y_2 \\ 
    &= C \oplus (0\text{x}1616\ldots1616)
\end{align*}

Once $K_{10}$ is retrieved, an adversary can recover all the previous subkeys and the original secret AES key. In the following, we will show how the subkey $K_{9}$ can be recovered from $K_{10}$. Here, $W^0_{10}$, $W^1_{10}$, $W^2_{10}$ and $W^3_{10}$ are known as the value of $K_{10}$ has been evaluated from the ciphertext previously using activating the hardware Trojan. 

\begin{enumerate}
    \item \textit{Step 1}: Computation of $W^3_{9}$ can be performed from XORing the  $W^3_{10}$ with $W^2_{10}$ as the XOR operation is reciprocal. 
    \begin{equation*}
        W^3_9 = W^3_{10} \oplus W^2_{10}
    \end{equation*}
    
    \item \textit{Step 2}: Once $W^3_9$ is known, one can compute $W^0_9$ using the following equation.
    
    \begin{equation*}
        W^0_9 = W^0_{10} \oplus f(W^3_{9})
    \end{equation*}
    
    \item \textit{Step 3}: Finally, $W^1_9$ and $W^2_9$ can be evaluated using the following equations.
    \begin{align*}
        W^1_9 &= W^1_{10} \oplus W^0_{10};
        &W^2_9 &= W^2_{10} \oplus W^1_{10}
    \end{align*}
    
\end{enumerate}

\newcommand{\conc}{\mathbin{\|}}
\begin{algorithm}[ht]
\SetAlgoLined
\caption{Reverse Key Schedule} \label{alg:inv_key_schedule}
\SetKwInput{KwInput}{Input}                
\SetKwInput{KwOutput}{Output}              
  
  \KwInput{SubKey $(K_{10})$ of round $R_{10}$}
  \KwOutput{Original Key~$(K)$}
\BlankLine
\For{$i =10$ \KwTo $1$ }{
        $[W^0_i, W^1_i,W^2_i,W^3_i] \gets assign(K_i)$ \;
        $W^3_{i-1} \gets W^3_i \oplus W^2_i$ \;
        $W^2_{i-1} \gets W^2_i \oplus W^1_i$ \;
        $W^1_{i-1} \gets W^1_i \oplus W^0_i$ \;
        $W^0_{i-1} \gets W^0_i \oplus f(W^3_{i-1})$ \;
    }
    $K \gets \{W^0_0 \conc W^1_0 \conc W^2_0 \conc W^3_0 \}$ \;
    Report $K$ \;
    
  \SetKwFunction{FMain}{$f$}
  \SetKwFunction{FAssign}{$assign$}

 \BlankLine
  \SetKwProg{Fn}{Function}{:}{}
  \Fn{\FAssign{$K_i$}}{
        $[k^1_i, k^2_i \dots k^{16}_i] \gets K_i$ \;
        $W^0_i \gets \{k_i^0~\conc~ k_i^1~\conc~k_i^2~\conc~k_i^3 \}$ \;  
         $W^1_i \gets \{k_i^4~\conc~k_i^5~\conc~k_i^6~\conc~k_i^7 \}$ \;
         $W^2_i \gets \{k_i^8~\conc~k_i^9~\conc~k_i^{10}~\conc k_i^{11} \}$ \;
          $W^3_i \gets \{k_i^{12}~\conc~k_i^{13}~\conc~k_i^{14}~\conc~k_i^{15} \}$ \;
        Return $[W^0_i, W^1_i,W^2_i,W^3_i]$ \;
  }

\BlankLine  
  \SetKwProg{Fn}{Function}{:}{}
  \Fn{\FMain{$W$}}{ 
     $G^0 \gets S(k^{13}) \oplus RC_i $ \;
     $G^1 \gets S(k^{14})$  \;
     $G^2 \gets S(k^{15})$ \;
     $G^3 \gets S(k^{12})$ \;
     Return $[G^0, G^1, G^2, G^3]$ \;
  }
\end{algorithm}

The general process for evaluating the secret key $K$ from $K_{10}$ is described in Algorithm~\ref{alg:inv_key_schedule}. The round subkey $K_{10}$ is provided as input to the algorithm and the original key $K$ will be returned as the output. The algorithm starts by selecting the subkey from which the previous round subkey is to be calculated~(Line 1). The subkey~($K_i$) is divided into 4 subgroups of 32-bits each, namely $[W^0_i, W^1_i,W^2_i,W^3_i]$, from the 128-bit key using the \textit{assign} function (Line 2). The 4 subgroups for $(i-1)^{th}$ key is calculated from the $i^{th}$ key (refer Figure~\ref{fig:key_schedule} in Section~\ref{sec:background})~(Lines 3-6). Finally, 4 different 32-bit subgroups~(\ie $W^0_0, W^1_0, W^2_0, W^3_0$) are obtained for the original key which are concatenated together and the algorithm reports the original key~$K$ (Lines 8-9). During these operations, function $f$ is used which takes inputs as the 32-bit subgroup word~$W$ (Line 17). Function $f$ performs the SubByte operation on 8-bits of keys using the sbox~($S$) and $RC_i$ (corresponding to each round) and returns the concatenated result~(Lines 18-22).  

\subsection{Tampering Attack on OpenCores AES Benchmark} The utilization of the hardware resources can be reduced by adopting multicycle designs, that re-use the hardware or functional blocks in a design. For the OpenCores AES~\cite{opencores} implementation, the plaintext~($P$) is provided as the input, and the internal result after every round is stored in round registers, which is then fed back to the SubByte layer of the next round. The result in the round register after $10^{th}$ round~($R_{10}$) is the ciphertext~($C$) output of the AES core. 

\begin{figure}[t]
    \centering
    \includegraphics{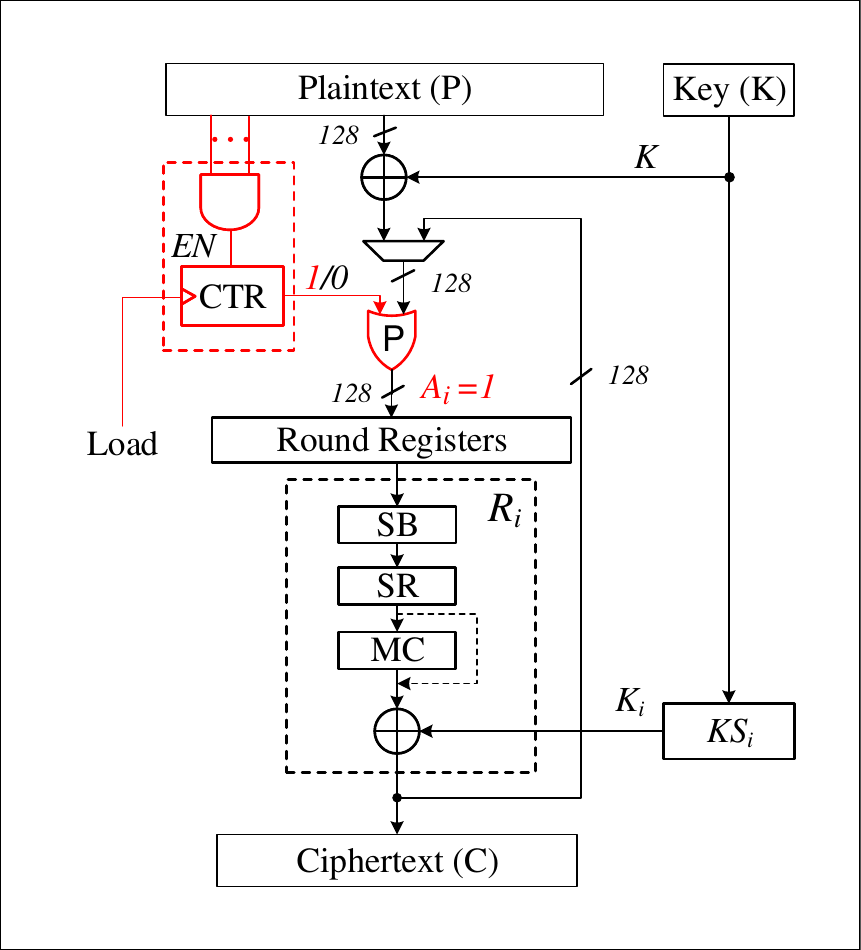}
    \caption{Tampering attack on OpenCores AES benchmark with a sequential hardware Trojan.} \vspace{-10px}
    \label{fig:multi_AES}
\end{figure}

Figure~\ref{fig:multi_AES} shows the hardware implementation of a OpenCores AES benchmark. The design has a \textit{load} input pin which loads the input plaintext for encryption. The total encryption process takes 13 clock cycles once the \textit{start} signal is assigned and the ciphertext is observed at the output when done signal becomes 1~\cite{opencores}. The proposed tampering attack can also be extended to this OpenCores AES design, which is tampered with a sequential Trojan described in section~\ref{sec:sequential_Trojan}. The counter~(CTR) of the hardware Trojan uses the \textit{load} signal as a clock. In other words, the Trojan counter will increase its value when the \textit{load} signal is high as well as the plaintext matches the trigger input pattern~(\ie $EN = 1$). It is necessary to trigger the Trojan $Q$ times consecutively to launch the attack. Once activated, the payload is delivered to make $A_i$ all 1's. The Trojan will remain activated during the entire encryption process and all the internal round computation will be modified as well. The ciphertext~($C$) observed at the output will be used to determine $K_{10}$, which can be computed as: $K_{10} = C \oplus (0\text{x}1616\ldots1616)$. Finally, the original key $(K)$ can be retrieved using the Algorithm~\ref{alg:inv_key_schedule}. Note that the attacks on AES are explained using a sequential Trojan. One can use other types of existing hardware Trojan designs to launch the attack as well. 

\section{Results and Discussions} \label{sec:experimentation}

To validate the effectiveness of our proposed attack, we implemented our proposed hardware Trojan in the OpenCores AES benchmark~\cite{opencores}. The sequential Trojan Trigger comprised of a AND gate followed by a counter. The Trigger inputs to the AND gate were taken directly from the plaintext input to the AES core. The trigger pattern was selected in such a way so that it does not belong to the manufacturing test patterns~(\eg stuck-at fault patterns)~\cite{jain2019taal,zhou2018modeling,jain2020special}.

\begin{table}[ht]
\centering \vspace{-5px}
\caption{Area overhead analysis.} \label{tab:area-overhead} 
\centering \vspace{-5px}
\begin{tabular}{|c|c|c|c|c|c|c|}
\hline
\multicolumn{1}{|c|}{Max Count (Q)} & 2  & 4 & 8 & 16 & 32 & 64   \\ \hline
Area Overhead (\%)  & 0.51 & 0.53 & 0.55 & 0.58 & 0.60 & 0.63 \\ \hline
\end{tabular}
\end{table}

The area for a hardware Trojan can vary based on the trigger input, trigger design, and the type of Trojan selected. For our experimentation, we implemented a sequential hardware Trojan with 5-input AND gate, counter with maximum count Q, and 128 payload OR gates. To estimate the area and power overhead, the Trojan inserted AES benchmark is synthesized with 32nm technology~\cite{32nmlibrary} using Synopsys Design Compiler~\cite{SynopsysDC}. Table~\ref{tab:area-overhead} shows the percentage area overhead obtained by comparing the Trojan-free and Trojan inserted AES benchmark. The overall area overhead for counter with different maximum count value $Q$. The higher value of $Q$ increases the difficulty of detecting a hardware Trojan using a logic test, as it is increasingly difficult to trigger the Trojan $Q$ times consecutively. The overhead is minimal and is less than 1\%. For example, it is only 0.55\%, when we choose $Q$ of 8. Note that the majority of the overhead comes from the payload as we require 128 OR gates. Since the Trojan remains quiet during normal operation, it does not have any switching power. The leakage power for the Trojan would only contribute to the power overhead. However, the Trojan's leakage power is in order of magnitude less than the switching power of the counter and can be negligible.


\section{Conclusion} \label{sec:conclusion}
Hardware Trojans can pose a severe threat to our critical infrastructure that relies on AES for encrypting sensitive data. We presented a novel tampering attack on AES core to extract the secret key by implanting a sequential hardware Trojan. The attack mainly relies on modifying the input for an internal round using the Trojan's payload, to mask the previous round's information. The Trojan helps an adversary to compute the last round subkey from the observed ciphertext. We present an algorithm to retrieve the original key from the last round's subkey. The sequential Trojan presented in the paper requires triggering of $Q$ consecutive times, which fulfills the requirement of increased difficulty in detecting such Trojans. It is extremely difficult to identify such Trojans using logic testing as the attacker only knows the trigger condition derived from the input plaintext, and apply it repeatedly.

\section*{Acknowledgment} 
This work was supported in parts by the National Science Foundation (NSF) under grant CNS-1755733 and Air Force Research Laboratory (AFRL) under grant AF-FA8650-19-1-1707. Any opinions, findings, and conclusions or recommendations expressed in this material are those of the authors and do not necessarily reflect the views of the NSF and AFRL.

\bibliographystyle{IEEEtran}
\bibliography{main}

\end{document}